\documentclass[aps, prd, showkeys, showpacs ]{revtex4}

\usepackage{amsmath,amsfonts,amscd,amssymb,epsf, multirow}
\usepackage[small]{caption}

\newcommand{\vep}{\varepsilon}

\begin{document}

\title{Exact classical sphere-plate  Casimir interaction   in $\boldsymbol{(D+1)}$-dimensional spacetime}

\author{L. P. Teo}
 \email{LeePeng.Teo@nottingham.edu.my}
 \affiliation{Department of Applied Mathematics, Faculty of Engineering, University of Nottingham Malaysia Campus, Jalan Broga, 43500, Semenyih, Selangor Darul Ehsan, Malaysia.}
\begin{abstract}
We consider the high temperature limit of the Casimir interaction between a Dirichlet sphere and a Dirichlet plate due to the vacuum fluctuations of a scalar field in $(D+1)$-dimensional Minkowski spacetime. The high temperature leading term of the Casimir free interaction energy is known as the classical term since it does not depend on the Planck constant $\hbar$. From the functional representation of the zero temperature Casimir interaction energy, we use Matsubara formalism to derive the classical term. It can be expressed as a weighted sum over logarithms of determinants. Using similarity transforms of matrices, we re-express this classical term as an infinite series. This series is then computed exactly using generalized Abel-Plana summation formula. From this, we deduce the short distance asymptotic expansions of the classical Casimir interaction force. As expected, the leading term agrees with the proximity force approximation. The next two terms in the asymptotic expansion are also computed. It is observed that the ratio of the next-to-leading order term to the leading order term is proportional to the dimension of spacetime. Hence, a larger correction to the proximity force approximation is expected in spacetime with higher dimensions. This is similar to a previous result deduced for the zero temperature case.
\end{abstract}

\maketitle
\section{Introduction}

Casimir interactions between objects of nontrivial geometries have been under active studies in the past ten years. This is partly motivated by the advent of nanotechnology which explores physics and technology in a length scale that renders Casimir interaction highly non-negligible. Another motivation comes from Casimir experiments where Casimir force is usually measured for the sphere-plate configuration due to the absence of alignment problem \cite{38}.

In the past few years, multiple scattering formalism has been used to cook up a recipe for computing the exact functional representation for the  Casimir interaction energy between any two objects \cite{6,7,8,9,10,11,12,13,14,15,16,17,18,19,20,21} in $(3+1)$-dimensional spacetime. In principle, one has to compute the scattering matrices of the objects in specific coordinate systems and the translation matrices between different coordinate frames. For objects with additional symmetries such as planes, cylinders and spheres, there are special coordinate systems available and the problem is tractable. Given the explicit formula for the Casimir interaction energy, one can then explore its properties numerically or analytically. Of particular interest is the small separation and large separation limits. The computation of large separation limit is usually straightforward. In the small separation regime which is of more concern to nanotechnology, it has been long believed that the leading Casimir interaction agrees with the proximity force approximation, and this has been verified for various geometric configurations \cite{15,16,22,23,24,25,26,27,28, 29, 30, 31}. However, to better reflect the actual strength of the Casimir force and for comparison to Casimir measurements, there is a need to go beyond proximity force approximation. The computation of the next-to-leading order term in the small separation asymptotic expansion is not an easy task \cite{15,16, 22, 23, 24, 25,26, 27, 28, 29, 30, 31}. A scheme based on derivative expansion has been proposed \cite{32,33} but yet to be verified.

Most of the above mentioned works only dealt with the zero temperature interaction. Nonetheless, the finite temperature effect cannot be neglected. Of particular appeal is the limit of the Casimir interaction in the high temperature regime. It has been long known that the high temperature leading term is linear in the temperature, given by the term  with zero Matsubara frequency. This term is known as the classical term since it does not depend on the Planck constant $\hbar$. In the small separation regime, except for the leading term which agrees with proximity force approximation, the asymptotic expansion of this classical term  is not much known. However, it was shown in \cite{4} that the classical Casimir interaction between a sphere and a plate  with Dirichlet boundary conditions can be computed exactly which   can be used to derive the full small separation asymptotic expansion.

Studying physics in higher dimensional spacetime has become a norm rather than an exception. Finite temperature Casimir effect inside a  rectangular cavity in $(D+1)$-dimensional Minkowski spacetime has been explored in \cite{34} more than 30 years ago.  As a first step to understand the Casimir effect between two nontrivial objects in higher dimensional spacetime, we studied the zero temperature Casimir effect between a sphere and a plate in $(D+1)$-dimensional Minkowski spacetime in \cite{1}. In this work, we extend the work to finite temperature regime and consider the high temperature limit. As in the three dimensional case considered in \cite{4}, we expect that the classical term can be computed exactly. Unlike \cite{4}, we do not make use of bi-spherical coordinates but use similarity transformation of matrices directly to obtain the result. An important tool in our computation is the generalized Abel-Plana summation formula \cite{35,36,37}.  From the exact formula for the classical Casimir interaction, we derive the small separation asymptotic expansion. In the case $D=3$, we recover the result obtained in \cite{4}.

\section{The Casimir free energy between a sphere and a plate}

In \cite{1}, we showed that when $D\geq 4$, the zero temperature Casimir interaction energy between a Dirichlet sphere of radius $R$ and a Dirichlet plate    can be  written as
\begin{align}\label{eq12_3_14}
E_{\text{Cas}}^{T= 0}=\frac{\hbar c}{2\pi}\int_0^{\infty} d \kappa \sum_{m=0}^{\infty}\frac{(2m+D-3)(m+D-4)!}{(D-3)!m!}\text{Tr}_m\,\ln\left(\mathbb{I}-\mathbb{M}_{m}(\kappa)\right),
\end{align}
where the elements $M_{m;l,l'}$ of $\mathbb{M}_{m}$ is
\begin{equation}\begin{split}
M_{m;l,l'}=&(-1)^{l+l'}  2^{2m+D-3} \Gamma\left(m+\frac{D-2}{2}\right)^2
\sqrt{\frac{\left(l+\frac{D-2}{2}\right)\left(l'+\frac{D-2}{2}\right)(l-m)!(l'-m)!}{(l+m+D-3)!(l'+m+D-3)!}}
\frac{I_{l+\frac{D-2}{2}}(\kappa R)}{K_{l+\frac{D-2}{2}}(\kappa  R)} \\&\times\int_{0}^{\infty}d\theta
 \left(\sinh\theta\right)^{2m +D-2}
C_{l-m}^{m+\frac{D-2}{2}}\left(\cosh\theta\right)C_{l'-m}^{m+\frac{D-2}{2}}\left(\cosh\theta\right)e^{-2\kappa L\cosh\theta}.\end{split}\end{equation}
Here $L$ is the distance from the center of the sphere to the plate. For fixed $m$, the trace Tr$_m$   is
\begin{align*}
\sum_{l=m}^{\infty}.
\end{align*}

When $D=3$, we can also represent the Casimir interaction energy by \eqref{eq12_3_14} provided that the summation $\sum_{m=0}^{\infty}$ is replaced by the summation $\sum_{m=0}^{\infty}\!'$, where the prime $\prime$ indicates that the term $m=0$ is summed with weight $1/2$.

Using Matsubara formalism, the finite temperature Casimir free interaction energy between a Dirichlet sphere and a Dirichlet plate can be obtained by replacing the integration over
$ \kappa  $ by summation over
$$\kappa_p=\frac{2\pi p k_BT}{\hbar c},$$
for $p$ from $-\infty$ to $\infty$. Namely,
\begin{align}\label{eq3_20_1}
E_{\text{Cas}}=k_BT\sum_{p=0}^{\infty}\!'\sum_{m=0}^{\infty}\frac{(2m+D-3)(m+D-4)!}{(D-3)!m!}\text{Tr}_m\,\ln\left(1-\mathbb{M}_{m}(\kappa_p)\right).
\end{align}

When $\kappa$ is large, $M_{m;l,l'}(\kappa)$ decays exponentially. Hence, in the high temperature regime, the contribution to the Casimir free interaction energy from those terms with $p\neq 0$ in \eqref{eq3_20_1} is exponentially small. The   high temperature limit of the free energy is given by the $p=0$ term in \eqref{eq3_20_1}, which is called the classical term since it is independent of the Planck constant $\hbar$. Namely,
\begin{align}\label{eq3_24_3}
E_{\text{Cas}}^{\text{classical}}=\frac{k_BT}{2} \sum_{m=0}^{\infty}\frac{(2m+D-3)(m+D-4)!}{(D-3)!m!}\lim_{\kappa\rightarrow 0}\text{Tr}\,\ln\left(1-\mathbb{M}_{m}(\kappa)\right).
\end{align}Here we do not put directly  $\kappa=0$ since $M_{m;l,l'}(\kappa)$ might not be well-defined when $\kappa=0$.
 Nevertheless,    the limit $$\lim_{\kappa\rightarrow 0}\text{Tr}\,\ln\left(1-\mathbb{M}_{m}(\kappa)\right)$$
should be well-defined and this is what we are going to compute.

As $z\rightarrow 0$,
\begin{align*}
I_{\nu}(z)\sim & \frac{1}{\Gamma(\nu+1)}\left(\frac{z}{2}\right)^{\nu},\\
K_{\nu}(z)\sim &\frac{\Gamma(\nu)}{2}\left(\frac{z}{2}\right)^{-\nu}.
\end{align*}Hence,
we find that as $\kappa\rightarrow 0$,
\begin{equation}\label{eq3_25_1}
\frac{I_{l+\frac{D-2}{2}}(\kappa R)}{K_{l+\frac{D-2}{2}}(\kappa  R)}\sim \frac{1}{2^{2l+D-3}\Gamma\left(l+\frac{D-2}{2}\right)\Gamma\left(l+\frac{D}{2}\right)}\left( \kappa R \right)^{2l+D-2}.
\end{equation}
By making a change of variables $u=\kappa\cosh\theta$, we have
\begin{equation}\label{eq3_24_4}
\begin{split}
&\int_{0}^{\infty}d\theta
 \left(\sinh\theta\right)^{2m +D-2}
C_{l-m}^{m+\frac{D-2}{2}}\left(\cosh\theta\right)C_{l'-m}^{m+\frac{D-2}{2}}\left(\cosh\theta\right)e^{-2\kappa L\cosh\theta}\\
=&\frac{1}{\kappa^{2m+D-2}}\int_{\kappa}^{\infty}du
 \left(u^2-\kappa^2\right)^{ m +\frac{D-3}{2}}
C_{l-m}^{m+\frac{D-2}{2}}\left(\frac{u}{\kappa}\right)C_{l'-m}^{m+\frac{D-2}{2}}\left(\frac{u}{\kappa}\right)e^{- 2Lu}.
\end{split}
\end{equation}To obtain the leading behavior of this integral when $\kappa\rightarrow 0$, we need to find the leading term of the Gegenbauer polynomials $C_n^{\nu}(z)$ when $z$ is large. From the Rodrigues' formula for the Gegenbauer polynomial \cite{2,3}:
\begin{align*}
C_n^{\nu}(z)=\frac{1}{2^n}\frac{\Gamma(2\nu+n)\Gamma\left(\nu+\frac{1}{2}\right)}{\Gamma(2\nu)\Gamma\left(\nu+\frac{1}{2}+n\right)}\frac{(z^2-1)^{\frac{1}{2}-\nu}}{n!}
\frac{d^n}{dz^n}\left(z^2-1\right)^{n+\nu-\frac{1}{2}},
\end{align*}we find that the leading term of the polynomial $C_n^{\nu}(z)$ is
\begin{align*}
C_n^{\nu}(z) =\frac{1}{2^nn!}\frac{\Gamma(2\nu+2n)\Gamma\left(\nu+\frac{1}{2}\right)}{\Gamma(2\nu)\Gamma\left(\nu+\frac{1}{2}+n\right)}z^n+\ldots.
\end{align*}
Hence, as $\kappa\rightarrow 0$, the leading term of the integral \eqref{eq3_24_4} is given by
\begin{equation}\label{eq3_25_2}
\begin{split}
&\frac{1}{\kappa^{l+l'+D-2}}\frac{1}{2^{l+l'-2m}(l-m)!(l'-m)!}
\frac{\Gamma(2l+D-2)\Gamma\left(2l'+D-2\right)\Gamma\left(m+\frac{D-1}{2}\right)^2}{\Gamma\left(2m+D-2\right)^2\Gamma\left(l+\frac{D-1}{2}\right)
\Gamma\left(l'+\frac{D-1}{2}\right)}\int_{0}^{\infty}du u^{ l+l'+D-3}e^{- 2Lu}\\
=&\frac{1}{\kappa^{l+l'+D-2}}\frac{1}{2^{2l+2l'-2m+D-2}(l-m)!(l'-m)!}
\frac{\Gamma(2l+D-2)\Gamma\left(2l'+D-2\right)\Gamma\left(m+\frac{D-1}{2}\right)^2}{\Gamma\left(2m+D-2\right)^2\Gamma\left(l+\frac{D-1}{2}\right)
\Gamma\left(l'+\frac{D-1}{2}\right)}\frac{\Gamma\left(l+l'+D-2\right)}{L^{l+l'+D-2}}\\
=&\frac{1}{\kappa^{l+l'+D-2}}\frac{1}{2^{2m+D-2}(l-m)!(l'-m)!}
\frac{\Gamma\left(l+\frac{D-2}{2}\right)\Gamma\left(l'+\frac{D-2}{2}\right)}{\Gamma\left(m+\frac{D-2}{2}\right)^2}\frac{\Gamma\left(l+l'+D-2\right)}{L^{l+l'+D-2}}.
\end{split}
\end{equation}
In the last row, we have used the identity
\begin{align*}
\Gamma(2z)=\frac{2^{2z-1}}{\sqrt{\pi}}\Gamma(z)\Gamma\left(z+\frac{1}{2}\right).
\end{align*}
From \eqref{eq3_25_1} and \eqref{eq3_25_2}, one can deduce that as $\kappa\rightarrow 0$,
\begin{align*}
M_{m;l,l'}(\kappa)\sim \kappa^{l-l'}.
\end{align*}
Since the trace of a matrix is not changed if the matrix is replaced by a similar matrix, we define
\begin{align*}
\widetilde{\mathbb{M}}_m=\mathbb{P}_1^{-1}\mathbb{M}_m\mathbb{P}_1,
\end{align*}
where $\mathbb{P}_1$ is a diagonal matrix with elements
\begin{align*}
\mathbb{P}_{1;l,l'}=(-1)^l\sqrt{\frac{(l+m+D-3)!}{\left(l+\frac{D-2}{2}\right)(l-m)!}}\frac{1}{\Gamma\left(l+\frac{D-2}{2}\right)}\left(\frac{\kappa R}{2}\right)^{l}\delta_{l,l'}.
\end{align*}
Then
\begin{align*}
\widetilde{M}_{m;l,l'}(\kappa) =(-1)^{-l+l'}\sqrt{\frac{\left(l+\frac{D-2}{2}\right)(l'+m+D-3)!(l-m)!}{\left(l'+\frac{D-2}{2}\right)(l+m+D-3)!(l'-m)!}}\frac{\Gamma\left(l+\frac{D-2}{2}\right)}{\Gamma\left(l'+\frac{D-2}{2}\right)}\left(\frac{\kappa R}{2}\right)^{-l+l'}M_{m;l,l'}(\kappa),
\end{align*}
and it follows that
\begin{align*}
\widetilde{M}_{m;l,l'}(0) =&\frac{(l+l'+D-3)!}{\left(l+m+D-3\right)!(l'-m)!}\left(\frac{R}{2L}\right)^{l+l'+D-2}.
\end{align*}
Hence, the classical Casimir interaction energy is
\begin{align}\label{eq3_26_1}
E_{\text{Cas}}^{\text{classical}}=\frac{k_BT}{2} \sum_{m=0}^{\infty}\frac{(2m+D-3)(m+D-4)!}{(D-3)!m!}  \ln\det\left(\mathbb{I}-\mathbb{N}_{m}\right),
\end{align}
where
\begin{align*}
N_{m;l,l'}=\frac{(l+l'+D-3)!}{\left(l+m+D-3\right)!(l'-m)!}\left(\frac{R}{2L}\right)^{l+l'+D-2},\hspace{1cm}l, l'\geq m.
\end{align*}
Since $L=R+d$, where $d$ is the distance from the sphere to the plate, the determinant $\det\left(\mathbb{I}-\mathbb{N}_{m}\right)$ is finite.
In the next section, we will derive an alternative expression for this determinant.

\section{Alternative expression for the classical Casimir interaction energy}

In this section, we use similarity transforms of matrices to find an alternative expression for the classical Casimir interaction energy \eqref{eq3_26_1}. Essentially, we transform  the matrices $\mathbb{N}_m$ to upper triangular matrices and use the fact that the determinant of an upper triangular matrix is equal to the product of its diagonal elements.

Let
\begin{equation}\label{eq3_26_3}x=\frac{R}{2L}=\frac{1}{2(1+\vep)},\hspace{1cm}\vep=\frac{d}{R}.\end{equation}
Let  $\mathbb{P}_2$ be a lower triangular matrix with elements
\begin{align*}
(P_{2})_{l,l'}=\left\{\begin{aligned}\frac{y^{l-l'}}{(l-l')!}\frac{(l-m)!}{(l'-m)!},\hspace{1cm} &l\geq l'\\
0,\hspace{3cm} &l<l'\end{aligned}\right.,
\end{align*}where $0<y<1$ is such that
\begin{equation}\label{eq3_26_4}y+y^{-1}=\frac{1}{x}.\end{equation}
One can check that the inverse   $\mathbb{P}_2^{-1}$ has elements
\begin{align*}
(P_2^{-1})_{l,l'}=\left\{\begin{aligned}(-1)^{l-l'}\frac{y^{l-l'}}{(l-l')!}\frac{(l-m)!}{(l'-m)!},\hspace{1cm} &l\geq l'\\
0,\hspace{4cm} &l<l'\end{aligned}\right..
\end{align*}
Using the fact that
\begin{align*}
\frac{1}{(1-v)^{n+1}}=\sum_{j=0}^{\infty}\frac{(n+j)!}{n!j!}v^j,
\end{align*}
we find that
\begin{align*}
\left(\mathbb{P}_2^{-1}\mathbb{N}_m\mathbb{P}_2\right)_{l,l'}=&\sum_{l_1=m}^l\sum_{l_2=l'}^{\infty}(-1)^{l-l_1}\frac{y^{l-l_1}}{(l-l_1)!}\frac{(l-m)!}{(l_1-m)!}
\frac{(l_1+l_2+D-3)!}{\left(l_1+m+D-3\right)!(l_2-m)!}x^{l_1+l_2+D-2}\frac{y^{l_2-l'}}{(l_2-l')!}\frac{(l_2-m)!}{(l'-m)!}\\
=&\sum_{l_1=m}^l\sum_{l_2=0}^{\infty}(-1)^{l-l_1}\frac{y^{l-l_1}}{(l-l_1)!}\frac{(l-m)!}{(l_1-m)!}
\frac{(l_1+l_2+l'+D-3)!}{\left(l_1+m+D-3\right)! }x^{l_1+l_2+l'+D-2}\frac{y^{l_2}}{l_2!}\frac{1}{(l'-m)!}\\
=&\sum_{l_1=m}^l (-1)^{l-l_1}\frac{y^{l-l_1}}{(l-l_1)!}\frac{(l-m)!}{(l_1-m)!}
\frac{(l_1 +l'+D-3)!}{\left(l_1+m+D-3\right)! }\left(\frac{x}{1-xy}\right)^{l_1+l'+D-2} \frac{1}{(l'-m)!}
\\
=&\sum_{l_1=m}^l (-1)^{l-l_1}\frac{y^{l+l'+D-2}}{(l-l_1)!}\frac{(l-m)!}{(l_1-m)!}
\frac{(l_1 +l'+D-3)!}{\left(l_1+m+D-3\right)! }  \frac{1}{(l'-m)!}\\
=&y^{l+l'+D-2}\times\;\text{coefficient of $v^{l+m+D-3}$ in }\;(1-v)^{l-m}(1-v)^{-l'+m-1}\\
=&\left\{\begin{aligned} \frac{y^{l+l'+D-2}}{(l'-l)!}\frac{( l'+m+D-3)!}{(l+m+D-3)!},\hspace{1cm} &l'\geq l\\
0,\hspace{4cm} &l'<l\end{aligned}\right..
\end{align*}
Notice that $\mathbb{P}_2^{-1}\mathbb{N}_m\mathbb{P}_2$ is an upper triangular matrix, and the diagonal elements are
$$\left(\mathbb{P}_2^{-1}\mathbb{N}_m\mathbb{P}_2\right)_{l,l}=y^{2l+D-2}.$$
Hence,
\begin{align}\label{eq3_26_2}
\det\left(\mathbb{I}-\mathbb{N}_m\right)=\sum_{l=m}^{\infty}\ln\left(1-y^{2l+D-2}\right).
\end{align}

In fact, using a suitable matrix $\mathbb{P}_3$, one can transform $\mathbb{P}_2^{-1}\mathbb{N}_m\mathbb{P}_2$ into a diagonal matrix. We leave it to the reader to check that if   $\mathbb{P}_3$ is the upper triangular  matrix with elements
\begin{align*}
\left(P_{3}\right)_{l,l'}=\left\{\begin{aligned}\frac{z^{l-l'}}{(l'-l)!}\frac{(l'+m+D-3)!}{(l+m+D-3)!},\hspace{1cm} &l'\geq l\\
0,\hspace{3cm} &l'<l\end{aligned}\right.,
\end{align*}
where
$$z=y-y^{-1},$$then
\begin{align*}
&\left(\mathbb{P}_3^{-1}\mathbb{P}_2^{-1}\mathbb{N}_m\mathbb{P}_2\mathbb{P}_3\right)_{l,l'} =\delta_{l,l'}y^{2l+D-2}.
\end{align*}

Returning back to the classical Casimir interaction energy, we obtain from \eqref{eq3_26_1} and \eqref{eq3_26_2} that
\begin{equation}\label{eq3_21_1}\begin{split}
E_{\text{Cas}}^{\text{classical}}=&\frac{k_BT}{2} \sum_{m=0}^{\infty}\frac{(2m+D-3)(m+D-4)!}{(D-3)!m!} \sum_{l=m}^{\infty} \ln\left(1-y^{2l+D-2}\right)\\
=&\frac{k_BT}{2}\sum_{l=0}^{\infty} \sum_{m=0}^{l}\frac{(2m+D-3)(m+D-4)!}{(D-3)!m!}   \ln\left(1-y^{2l+D-2}\right)\\
=&\frac{k_BT}{2}\sum_{l=0}^{\infty}  \frac{(2l+D-2)(l+D-3)!}{(D-2)!l!}   \ln\left(1-y^{2l+D-2}\right).\end{split}\end{equation}
Notice that when $D=3$, \eqref{eq3_21_1} gives
\begin{align}\label{eq3_26_5}
E_{\text{Cas}}^{\text{classical}}=\frac{k_BT}{2}\sum_{l=0}^{\infty} (2l+1)   \ln\left(1-y^{2l+1}\right),
\end{align}
which is exactly the result derived in \cite{4} using bi-spherical coordinates.

From the definitions \eqref{eq3_26_4} and \eqref{eq3_26_3}, we find that
\begin{align*}
y=1+\vep-\sqrt{\vep^2+2\vep}=\frac{L}{R}-\sqrt{\left(\frac{L}{R}\right)^2-1}.
\end{align*}
When $L\gg R$,
\begin{align*}
y\sim \frac{1}{2}\frac{R}{L},
\end{align*}which shows that the large separation leading term of the classical Casimir interaction energy comes from the $l=0$ term in \eqref{eq3_21_1} and is given by
\begin{equation*}
E_{\text{Cas}}^{\text{classical}}\sim -\frac{k_BTR^{D-2}}{2^{D-1}L^{D-2}}.
\end{equation*}

\section{Samll separation exact formula of the classical Casimir interaction energy}

In this section, we use generalized Abel-Plana summation formula to compute the classical Casimir interaction energy \eqref{eq3_26_5}. We need to consider the case when $D$ is even and the case when $D$ is odd separately.

Let$$y=e^{-\mu},$$where
\begin{align}\label{eq3_26_7}\mu=-\ln\left(1+\vep-\sqrt{\vep^2+2\vep}\right)>0.\end{align}

First, we want to rewrite \eqref{eq3_21_1}.
When $D$ is even, let $$\tilde{l}=l+\frac{D-2}{2}.$$Then
\begin{align*}
E_{\text{Cas}}^{\text{classical}}=&\frac{k_BT}{2}\sum_{\tilde{l}=\frac{D-2}{2}}\frac{2}{(D-2)!}\tilde{l}\left(\tilde{l}+\frac{D-4}{2}\right)
\left(\tilde{l}+\frac{D-6}{2}\right)\ldots \left(\tilde{l}+1\right)\tilde{l}\left(\tilde{l}-1\right)\ldots\left(\tilde{l}-\frac{D-4}{2}\right)\ln\left(1-e^{-2\tilde{l}\mu}\right).\end{align*}
Notice that the summand is zero when $\tilde{l}=1, 2, \ldots, (D-4)/2$. This allows us to start the summation from $\tilde{l}=1$ instead of $(D-2)/2$. Moreover,
$$\tilde{l}\left(\tilde{l}+\frac{D-4}{2}\right)
\left(\tilde{l}+\frac{D-6}{2}\right)\ldots \left(\tilde{l}+1\right)\tilde{l}\left(\tilde{l}-1\right)\ldots\left(\tilde{l}-\frac{D-4}{2}\right)$$ is a polynomial of degree $D-2$ in $\tilde{l}$ which can be written as
\begin{align*}
\sum_{j=1}^{D-2}x_{D;j}\tilde{l}^j,
\end{align*}with $x_{D;j}=0$ when $j$ is odd.
Hence,
\begin{align*}
E_{\text{Cas}}^{\text{classical}}=&\frac{k_BT}{(D-2)!}\sum_{\tilde{l}=1}^{\infty}\sum_{j=1}^{D-2}x_{D;j}\tilde{l}^j\ln\left(1-e^{-2\tilde{l}\mu}\right)
\end{align*}

When $D$ is odd, let
$$\tilde{l}=l+\frac{D-3}{2}.$$We find in the same way that
\begin{align*}
E_{\text{Cas}}^{\text{classical}}=&\frac{k_BT}{2}\sum_{\tilde{l}=\frac{D-3}{2}}\frac{2}{(D-2)!}\left(\tilde{l}+\frac{1}{2}\right)\left(\tilde{l}+\frac{1}{2}+\frac{D-4}{2}\right)
\left(\tilde{l}+\frac{1}{2}+\frac{D-6}{2}\right)\ldots \left(\tilde{l}+\frac{1}{2}+\frac{1}{2}\right) \left(\tilde{l}+\frac{1}{2}-\frac{1}{2}\right)\ldots\\&\hspace{1cm}\times\left(\tilde{l}+\frac{1}{2}-\frac{D-4}{2}\right)\ln\left(1-e^{-\left(2\tilde{l}+1\right)\mu}\right)\\
=&\frac{k_BT}{(D-2)!}\sum_{\tilde{l}=0}^{\infty}\sum_{j=1}^{D-2}x_{D;j}\left(\tilde{l}+\frac{1}{2}\right)^j\ln\left(1-e^{-(2\tilde{l}+1)\mu}\right),
\end{align*}where now
\begin{align*}
\sum_{j=1}^{D-2}x_{D;j}\left(\tilde{l}+\frac{1}{2}\right)^j= \left(\tilde{l}+\frac{1}{2}\right)\left(\tilde{l}+\frac{1}{2}+\frac{D-4}{2}\right)
\left(\tilde{l}+\frac{1}{2}+\frac{D-6}{2}\right)\ldots \left(\tilde{l}+\frac{1}{2}+\frac{1}{2}\right) \left(\tilde{l}+\frac{1}{2}-\frac{1}{2}\right)\ldots \left(\tilde{l}+\frac{1}{2}-\frac{D-4}{2}\right),
\end{align*}with $x_{D;j}=0$ when $j$ is even.

As a function of the complex variable $z$, $\ln\left(1-e^{-az}\right)$ does not have good analytic properties. So instead of considering the classical Casimir energy, we consider the classical Casimir force.
Since
\begin{align*}
\mu'(\vep)=\frac{1}{\sqrt{\vep^2+2\vep}},
\end{align*}
we find that when $D$ is even,
\begin{align*}
F_{\text{Cas}}^{\text{classical}}=-\frac{k_BT}{R\sqrt{\vep^2+2\vep}}\frac{2}{(D-2)!}\sum_{\tilde{l}=0}^{\infty}\sum_{j=1}^{D-2}x_{D;j}\frac{\tilde{l}^{j+1}}{e^{2\tilde{l}\mu}-1};
\end{align*}
and when $D$ is odd,
\begin{align*}
F_{\text{Cas}}^{\text{classical}}=-\frac{k_BT}{R\sqrt{\vep^2+2\vep}}\frac{2}{(D-2)!}\sum_{\tilde{l}=0}^{\infty}\sum_{j=1}^{D-2}x_{D;j}\frac{\left(\tilde{l}+\frac{1}{2}\right)^{j+1}}{e^{\left(2\tilde{l}+1\right)\mu}-1}.
\end{align*} Notice that $x_{D;j}\neq 0$ only if $D$ and $j$ have the same parity.

Now we have to deal with functions of the form $$\frac{z^n}{e^{az}-1},$$ which is not analytic but is meromorphic. We cannot apply Abel-Plana summation formula, but instead we can apply the generalized Abel-Plana summation formula \cite{35,36,37}, which says that if $f(z)$ is a meromorphic function that only has poles on the imaginary axis,
\begin{equation}\label{eq7_27_1}
\begin{split}
\frac{1}{2}f(0)+\sum_{p=1}^{\infty}f(p)=&\int_0^{\infty} f(x) dx +i\int_0^{\infty}\frac{f(iy)-f(-iy)}{e^{2\pi y}-1}dy +\pi i \sum_{ y>0} \frac{\text{Res}_{z=iy} f(z)-\text{Res}_{z=-iy}f(z)}{ e^{2\pi y}-1},
\end{split}
\end{equation}
\begin{equation}\label{eq7_27_2}
\begin{split}
 \sum_{p=0}^{\infty}f(2p+1)=&\frac{1}{2}\int_0^{\infty} f(x) dx -\frac{i}{2}\int_0^{\infty}\frac{f(iy)-f(-iy)}{e^{\pi y}+1}dy -\frac{\pi i}{2} \sum_{ y>0} \frac{\text{Res}_{z=iy} f(z)-\text{Res}_{z=-iy}f(z)}{ e^{\pi y}+1}.
\end{split}
\end{equation}

When $D$ is even, we apply \eqref{eq7_27_1} with
\begin{align*}
f(z)=\frac{z^{j+1}}{e^{2 \mu z}-1},
\end{align*} where $j\geq 1$ is even.
$f(z)$ has poles at $z=\pm i\pi n/\mu$, $n=1, 2, \ldots$ and
\begin{align*}
\text{Res}_{z=\pm \frac{i\pi n}{\mu}}f(z)=\pm \frac{i^{j+1}\pi^{j+1}n^{j+1}}{2\mu^{j+2}}.
\end{align*}
Hence,
\begin{align*}
\pi i \sum_{ y>0} \frac{\text{Res}_{z=iy} f(z)-\text{Res}_{z=-iy}f(z)}{ e^{2\pi y}-1}=\frac{(-1)^{\frac{j}{2}+1}\pi^{j+2}}{\mu^{j+2}}\sum_{n=1}^{\infty} \frac{n^{j+1}}{e^{\frac{2\pi^2 n}{\mu}}-1}.
\end{align*}This sum goes to zero exponentially fast when $\mu\rightarrow0$.
On the other hand,
\begin{align*}
\int_0^{\infty}f(x)dx=&\int_0^{\infty}\frac{x^{j+1}}{e^{2\mu x}-1}dx\\
=&\frac{\Gamma(j+2)}{2^{j+2}\mu^{j+2}}\zeta(j+2),
\end{align*}
\begin{align*}
i\int_0^{\infty}\frac{f(iy)-f(-iy)}{e^{2\pi y}-1}dy=&(-1)^{\frac{j}{2}}\int_0^{\infty}\frac{y^{j+1}}{e^{2\pi y}-1}dy\\=&(-1)^{\frac{j}{2}}\frac{\Gamma(j+2)}{2^{j+2}\pi^{j+2}}\zeta(j+2).
\end{align*}Here $\zeta(s)=\sum_{n=1}^{\infty}1/n^s$ is the Riemann zeta function.
Since $f(0)=0$, the generalized Abel-Plana summation formula \eqref{eq7_27_1} implies that
\begin{align*}
\frac{\tilde{l}^{j+1}}{e^{2\tilde{l}\mu}-1}=&\frac{\Gamma(j+2)}{2^{j+2}\mu^{j+2}}\zeta(j+2)+(-1)^{\frac{j}{2}}\frac{\Gamma(j+2)}{2^{j+2}\pi^{j+2}}\zeta(j+2)+\frac{(-1)^{\frac{j}{2}+1}\pi^{j+2}}{\mu^{j+2}}\sum_{n=1}^{\infty} \frac{n^{j+1}}{e^{\frac{2\pi^2 n}{\mu}}-1}.
\end{align*}From this, we obtain the exact expression for the classical Casimir interaction force:
\begin{equation}\label{eq3_26_8}\begin{split}
F_{\text{Cas}}^{\text{classical}}=-\frac{k_BT}{R\sqrt{\vep^2+2\vep}}\frac{2}{(D-2)!} \sum_{j=1}^{D-2}x_{D;j}\left\{\frac{\Gamma(j+2)}{2^{j+2}\mu^{j+2}}\zeta(j+2)+(-1)^{\frac{j}{2}}\frac{\Gamma(j+2)}{2^{j+2}\pi^{j+2}}\zeta(j+2)
+\frac{(-1)^{\frac{j}{2}+1}\pi^{j+2}}{\mu^{j+2}}\sum_{n=1}^{\infty} \frac{n^{j+1}}{e^{\frac{2\pi^2 n}{\mu}}-1}\right\}.
\end{split}\end{equation}

When $D$ is odd,
we apply the generalized Abel-Plana summation formula \eqref{eq7_27_2} with
\begin{align*}
f(z)=\frac{1}{2^{j+1}}\frac{z^{j+1}}{e^{\mu z}-1},
\end{align*}
where $j\geq 1$ is odd.
Then
$f(z)$ has poles at $z=\pm 2i\pi n/\mu$ with
\begin{align*}
\text{Res}_{z=\pm \frac{2i\pi n}{\mu}}f(z)=  \frac{i^{j+1}\pi^{j+1}n^{j+1}}{\mu^{j+2}}.
\end{align*}
Consequently,
\begin{align*}
-\frac{\pi i}{2} \sum_{ y>0} \frac{\text{Res}_{z=iy} f(z)-\text{Res}_{z=-iy}f(z)}{ e^{\pi y}+1}=0.
\end{align*}
On the other hand,
\begin{align*}
\frac{1}{2}\int_0^{\infty}f(x)dx=\frac{1}{2^{j+2}}\int_0^{\infty} \frac{x^{j+1}}{e^{\mu x}-1}dx=\frac{\Gamma(j+2)}{2^{j+2}\mu^{j+2}}\zeta(j+2),
\end{align*}
\begin{align*}
f(iy)-f(-iy)=\frac{i^{j+1}}{2^{j+1}}\frac{y^{j+1}}{e^{i\mu y}-1}-\frac{i^{j+1}}{2^{j+1}}\frac{y^{j+1}}{e^{-i\mu y}-1}=\frac{i^j}{2^{j+1}}y^{j+1}\cot\frac{\mu y}{2},
\end{align*}
which gives
\begin{align*}
 -\frac{i}{2}\int_0^{\infty}\frac{f(iy)-f(-iy)}{e^{\pi y}+1}dy=\frac{(-1)^{\frac{j-1}{2}}}{2^{j+2}}\int_0^{\infty}\frac{y^{j+1}\cot\frac{\mu y}{2}}{e^{\pi y}+1}dy.
\end{align*}
From these, we obtain the exact formula for the classical Casimir interaction force:
\begin{equation}\label{eq3_26_9}\begin{split}
F_{\text{Cas}}^{\text{classical}}=-\frac{k_BT}{R\sqrt{\vep^2+2\vep}}\frac{2}{(D-2)!} \sum_{j=1}^{D-2}x_{D;j}\left\{\frac{\Gamma(j+2)}{2^{j+2}\mu^{j+2}}\zeta(j+2)+\frac{(-1)^{\frac{j-1}{2}}}{2^{j+2}}\int_0^{\infty}\frac{y^{j+1}\cot\frac{\mu y}{2}}{e^{\pi y}+1}dy\right\}.
\end{split}\end{equation}

From the definition of $\mu$ \eqref{eq3_26_7}, we find that as $\vep\ll 1$,
\begin{align*}
\mu\sim \sqrt{2\vep}.
\end{align*}
Hence, Eqs. \eqref{eq3_26_8} and \eqref{eq3_26_9} are ideal for studying the small separation asymptotic behavior of the Casimir interaction force. In particular, we find that when $\mu\ll 1$, 
\begin{equation}\label{eq3_27_3}\begin{split}
F_{\text{Cas}}^{\text{classical}}=-\frac{k_BT}{R\sqrt{\vep^2+2\vep}}\frac{2}{(D-2)!} \sum_{j=1}^{D-2}x_{D;j}\left\{\frac{\Gamma(j+2)}{2^{j+2}\mu^{j+2}}\zeta(j+2)+(-1)^{\frac{j}{2}}\frac{\Gamma(j+2)}{2^{j+2}\pi^{j+2}}\zeta(j+2)
 \right\}+O(\mu)
\end{split}\end{equation}if $D$ is even; 
and 
\begin{equation}\label{eq3_27_4}\begin{split}
F_{\text{Cas}}^{\text{classical}}=-\frac{k_BT}{R\sqrt{\vep^2+2\vep}}\frac{2}{(D-2)!} \sum_{j=1}^{D-2}x_{D;j}\left\{\frac{\Gamma(j+2)}{2^{j+2}\mu^{j+2}}\zeta(j+2)+\frac{(-1)^{\frac{j-1}{2}}}{2^{j+1}\mu}\int_0^{\infty}\frac{y^{j} }{e^{\pi y}+1}dy\right\}+O(\mu)
\end{split}\end{equation}
if $D$ is odd. In the latter, we have used the fact that
$$\cot\frac{\mu y}{2}\sim \frac{2}{\mu y}+O(\mu)$$ as $\mu\ll 1$.

\section{Comparison to proximity force approximation}
The proximity force approximation approximates the Casimir interaction force between two objects by summing  the local Casimir force density between two planes over the surfaces.
In $(D+1)$-dimensional Minkowski spacetime, the classical Casimir   force density between two parallel plates  both subject to Dirichlet boundary conditions is given by \cite{5}:
\begin{align*}
\mathcal{F}_{\text{Cas}}^{\text{classical}, \parallel} (d)=-k_BT\frac{(D-1)\Gamma\left(\frac{D}{2}\right)\zeta(D)}{2^{D}\pi^{\frac{D}{2}}}\frac{1}{d^{D}}=\frac{b_D}{d^{D}},
\end{align*}where $d$ is the distance between the two plates.

As in \cite{1},  we find that the proximity force approximation to the classical Casimir interaction force between a Dirichlet sphere and a Dirichlet plate in $(D+1)$-dimensional spacetime is
\begin{equation}\label{eq3_26_10}\begin{split}
F_{\text{Cas}}^{\text{classical}, \text{PFA}}=& R^{D-1}b_D\frac{2\pi^{\frac{D-1}{2}}}{\Gamma\left(\frac{D-1}{2}\right)}\int_0^{\pi}\frac{d\theta_1\sin^{D-2}\theta_1}{\left(d+R(1-\cos\theta_1)\right)^{D}}\\
\sim & b_D\frac{ \pi^{\frac{D}{2}}}{2^{\frac{D-1}{2}}\Gamma\left(\frac{D}{2}\right)}\frac{1}{R\vep^{\frac{D+1}{2}}}\\=&-k_BT\frac{(D-1)}{2^{\frac{3D-1}{2}}} \frac{\zeta(D)}{ R\vep^{\frac{D+1}{2}}}.
\end{split}\end{equation}

From \eqref{eq3_26_8} and \eqref{eq3_26_9}, we find that when $\vep\ll 1$, the leading term of the classical Casimir interaction force comes from the term with $j=D-2$. Since $x_{D;D-2}=1$, we have
\begin{equation*}\begin{split}
F_{\text{Cas}}^{\text{classical}}\sim &-\frac{k_BT}{R\sqrt{\vep^2+2\vep}}\frac{D-1}{2^{D-1}\mu^D}\zeta(D)\\
\sim &-k_BT\frac{(D-1)}{2^{\frac{3D-1}{2}}} \frac{\zeta(D)}{ R\vep^{\frac{D+1}{2}}},
\end{split}\end{equation*}which agrees with the proximity force approximation \eqref{eq3_26_10}.

\section{Small separation asymptotic expansion}
In this section, we derive the small separation asymptotic expansion of the classical Casimir interaction force from \eqref{eq3_26_8} and \eqref{eq3_26_9} in terms of $\vep=d/R$.

As $\vep\ll 1$,
\begin{align}\label{eq3_26_11}
\frac{1}{\sqrt{\vep^2+2\vep}}=&\frac{1}{\sqrt{2\vep}}\left(1-\frac{\vep}{4}+\frac{3}{32}\vep^2+\ldots\right),
\end{align}
\begin{align*}
\mu=&\sqrt{2\vep}\left(1-\frac{\vep}{12}+\frac{3\vep^2}{160}+\ldots\right),
\end{align*}
which gives
\begin{align}\label{eq3_26_12}
\frac{1}{\mu^D}=\frac{1}{2^{\frac{D}{2}}\vep^{\frac{D}{2}}}\left(1+\frac{D\vep}{12}+\frac{D(5D-22)}{1440}\vep^2 +\ldots\right).
\end{align}
For the constants $x_{D;j}$, it is straightforward to show that
\begin{equation}\label{eq3_27_11}\begin{split}
x_{D;D-2}=&1,\\
x_{D;D-3}=&0,\\
 x_{D;D-4}=&-\frac{(D-2)(D-3)(D-4)}{24},\\
 x_{D;D-5}=&0,\\
x_{D;D-6}=&\frac{(D-2)(D-3)(D-4)(D-5)(D-6)(5D-8)}{5760},\\
&\vdots
\end{split}\end{equation}
On the other hand,
\begin{align*}
\cot\frac{\mu y}{2}=\frac{2}{\mu y}-\frac{\mu y}{6}+\ldots.
\end{align*}
Hence,
\begin{equation}\label{eq3_26_13}\begin{split}
&\frac{(-1)^{\frac{j-1}{2}}}{2^{j+2}}\int_0^{\infty}\frac{y^{j+1}\cot\frac{\mu y}{2}}{e^{\pi y}+1}dy\\
=&\frac{(-1)^{\frac{j-1}{2}}}{2^{j+2}}\int_0^{\infty}\frac{y^{j+1}\left(\frac{2}{\mu y}-\frac{\mu y}{6}+\ldots\right)}{e^{\pi y}+1}dy\\
=&\frac{(-1)^{\frac{j-1}{2}}}{2^{j+1}}\frac{1}{\mu}\frac{\Gamma(j+1)}{\pi^{j+1}}(1-2^{-j})\zeta(j+1)
+\frac{(-1)^{\frac{j+1}{2}}}{3\times2^{j+3}} \mu\frac{\Gamma(j+3)}{\pi^{j+3}}(1-2^{-j-2})\zeta(j+3)+O(\mu^3).
\end{split}\end{equation}
When $D=3$, \eqref{eq3_26_9}, \eqref{eq3_26_13}, \eqref{eq3_26_11} and \eqref{eq3_26_12} give
\begin{equation}\label{eq3_27_5}\begin{split}
F_{\text{Cas}}^{\text{classical}}=&-\frac{2k_BT}{R\sqrt{\vep^2+2\vep}} \left\{\frac{1}{4\mu^{3}}\zeta(3)+\frac{1}{8}\int_0^{\infty}\frac{y^{4}\cot\frac{\mu y}{2}}{e^{\pi y}+1}dy\right\}\\
=&-\frac{2k_BT}{R\sqrt{2\vep}}\left(1-\frac{\vep}{4}+\frac{3}{32}\vep^2+\ldots\right)\\&\times\left(\frac{1}{8\sqrt{2}\vep^{\frac{3}{2}}}\zeta(3)\left(1+\frac{\vep}{4}-\frac{7\vep^2}{480}+\ldots\right)
+\frac{1}{48\sqrt{2 \vep}} \left(1+\frac{\vep}{12}+\ldots\right)-\frac{7}{5760}\sqrt{2\vep}+\ldots\right)\\
=&-\frac{k_BT}{8R\vep^2}\zeta(3)\left(1+\frac{1}{6\zeta(3)}\vep+\left(\frac{1}{60}-\frac{17}{360\zeta(3)}\right)\vep^2+\ldots\right).
\end{split}\end{equation}

\begin{figure}[h]
\epsfxsize=0.5\linewidth \epsffile{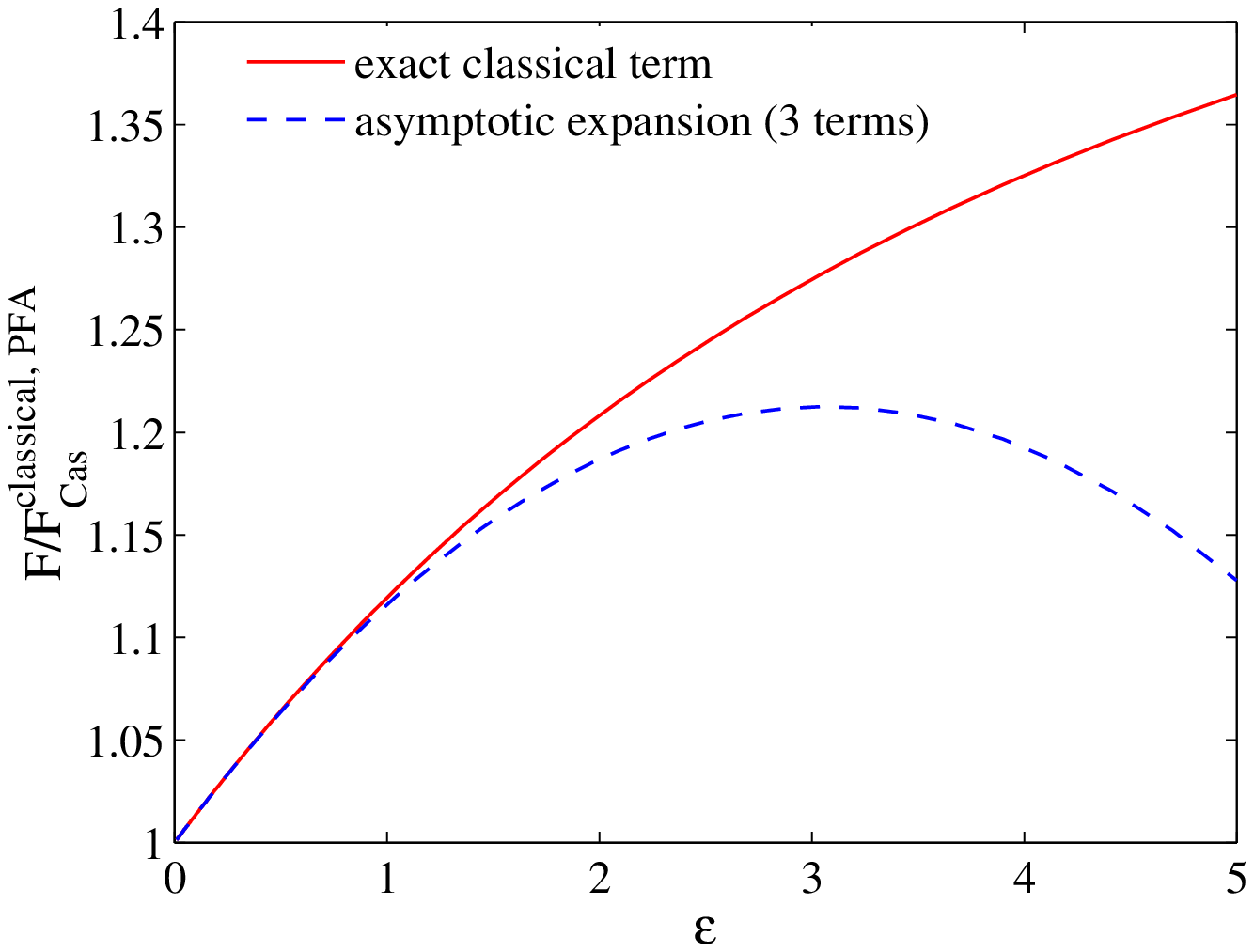} \caption{\label{f1} The comparison between the exact Casimir interaction force with the   asymptotic expansion \eqref{eq3_27_5} when $D=3$. Both quantities are normalized by the proximity force approximation. }\end{figure}

When $D=4$, \eqref{eq3_26_8}, \eqref{eq3_26_11} and \eqref{eq3_26_12} give
\begin{equation}\label{eq3_27_6}\begin{split}
F_{\text{Cas}}^{\text{classical}}= &-\frac{k_BT}{R\sqrt{\vep^2+2\vep}}  \left(\frac{3}{8\mu^{4}}\zeta(4)-\frac{3}{8\pi^{4}}\zeta(4)+\ldots\right)\\
=&-k_BT\frac{3\zeta(4)}{32\sqrt{2}R\vep^{\frac{5}{2}}}\left(1-\frac{\vep}{4}+\frac{3}{32}\vep^2+\ldots\right) \left(1+\frac{\vep}{3}-\frac{\vep^2}{180}-\frac{4\vep^2}{\pi^4}+\ldots\right)\\
=&-k_BT\frac{3\zeta(4)}{32\sqrt{2}R\vep^{\frac{5}{2}}}\left(1+\frac{\vep}{12}+\left(\frac{7}{1440}-\frac{4}{\pi^4}\right)\vep^2+\ldots\right).
\end{split}\end{equation}

\begin{figure}[h]
\epsfxsize=0.5\linewidth \epsffile{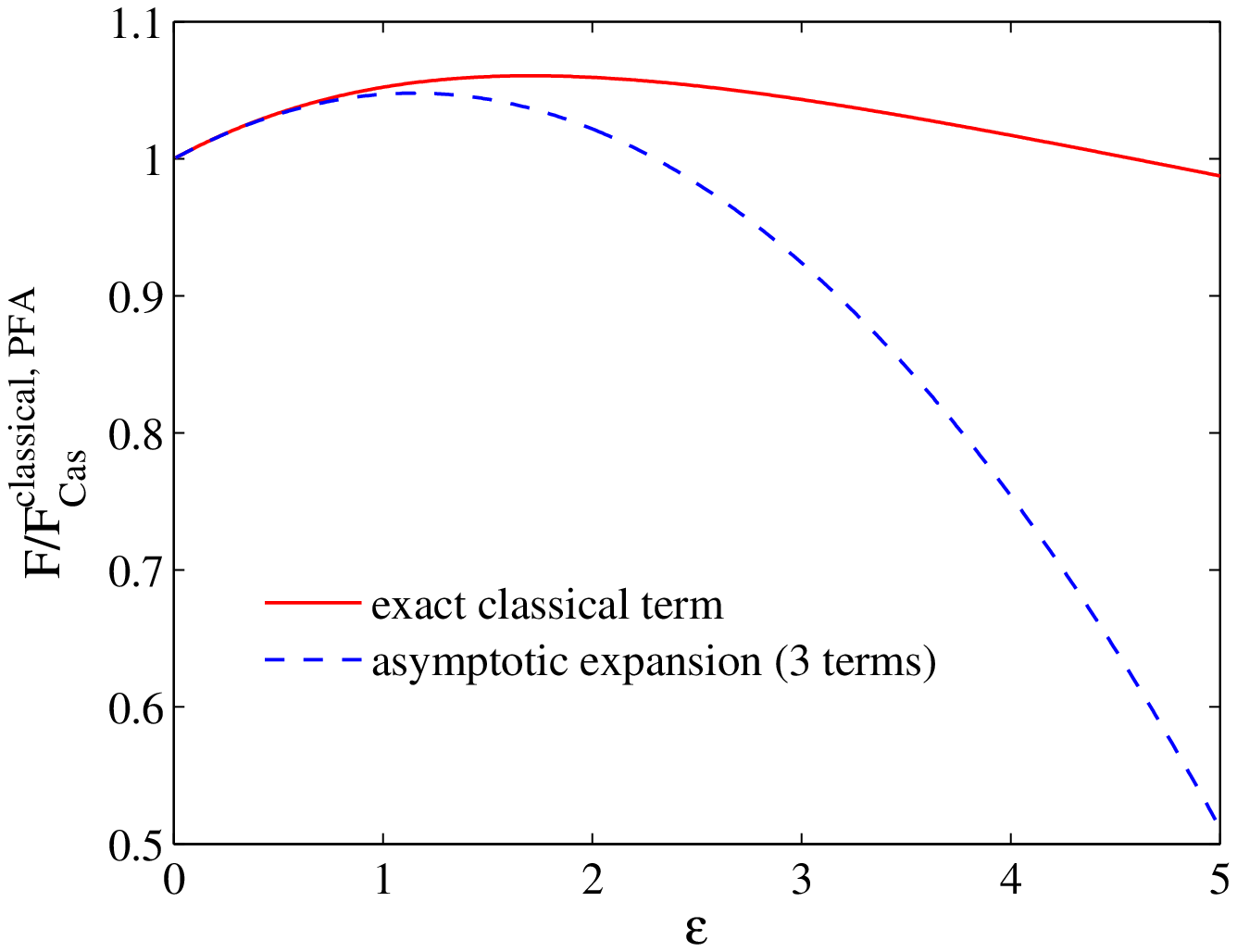} \caption{\label{f2} The comparison between the exact Casimir interaction force with the   asymptotic expansion \eqref{eq3_27_6} when $D=4$. Both quantities are normalized by the proximity force approximation. }\end{figure}

When $D=5$, \eqref{eq3_26_9}, \eqref{eq3_26_13}, \eqref{eq3_26_11} and \eqref{eq3_26_12} give
\begin{equation}\label{eq3_27_7}\begin{split}
F_{\text{Cas}}^{\text{classical}}=&-\frac{k_BT}{3R\sqrt{\vep^2+2\vep}}  \left\{\frac{3}{4\mu^{5}}\zeta(5)-\frac{1}{32}\int_0^{\infty}\frac{y^{4}\cot\frac{\mu y}{2}}{e^{\pi y}+1}dy-\frac{1}{4}\left(\frac{1}{4\mu^{3}}\zeta(3)+\frac{1}{8}\int_0^{\infty}\frac{y^{2}\cot\frac{\mu y}{2}}{e^{\pi y}+1}dy\right)\right\}\\
=&-\frac{k_BT}{3R\sqrt{2\vep} }\left(1-\frac{\vep}{4}+\frac{3}{32}\vep^2+\ldots\right)\\&\hspace{2cm}\times\left(\frac{3\zeta(5)}{16\sqrt{2}\vep^{\frac{5}{2}}}
\left(1+\frac{5\vep}{12}+\frac{\vep^2}{96}+\ldots\right)-\frac{7}{1920}\frac{1}{\sqrt{2\vep}}
-\frac{\zeta(3)}{32\sqrt{2}\vep^{\frac{3}{2}}}\left(1+\frac{\vep}{4}+\ldots\right)-\frac{1}{192}\frac{1}{\sqrt{2\vep}}+\ldots\right)\\
=&-\frac{k_BT}{32R\vep^3}\zeta(5)\left(1+\frac{\vep}{6}-\frac{\zeta(3)}{6\zeta(5)}\vep
-\frac{17}{360\zeta(5)}\vep^2+\ldots\right).
\end{split}\end{equation}

\begin{figure}[h]
\epsfxsize=0.5\linewidth \epsffile{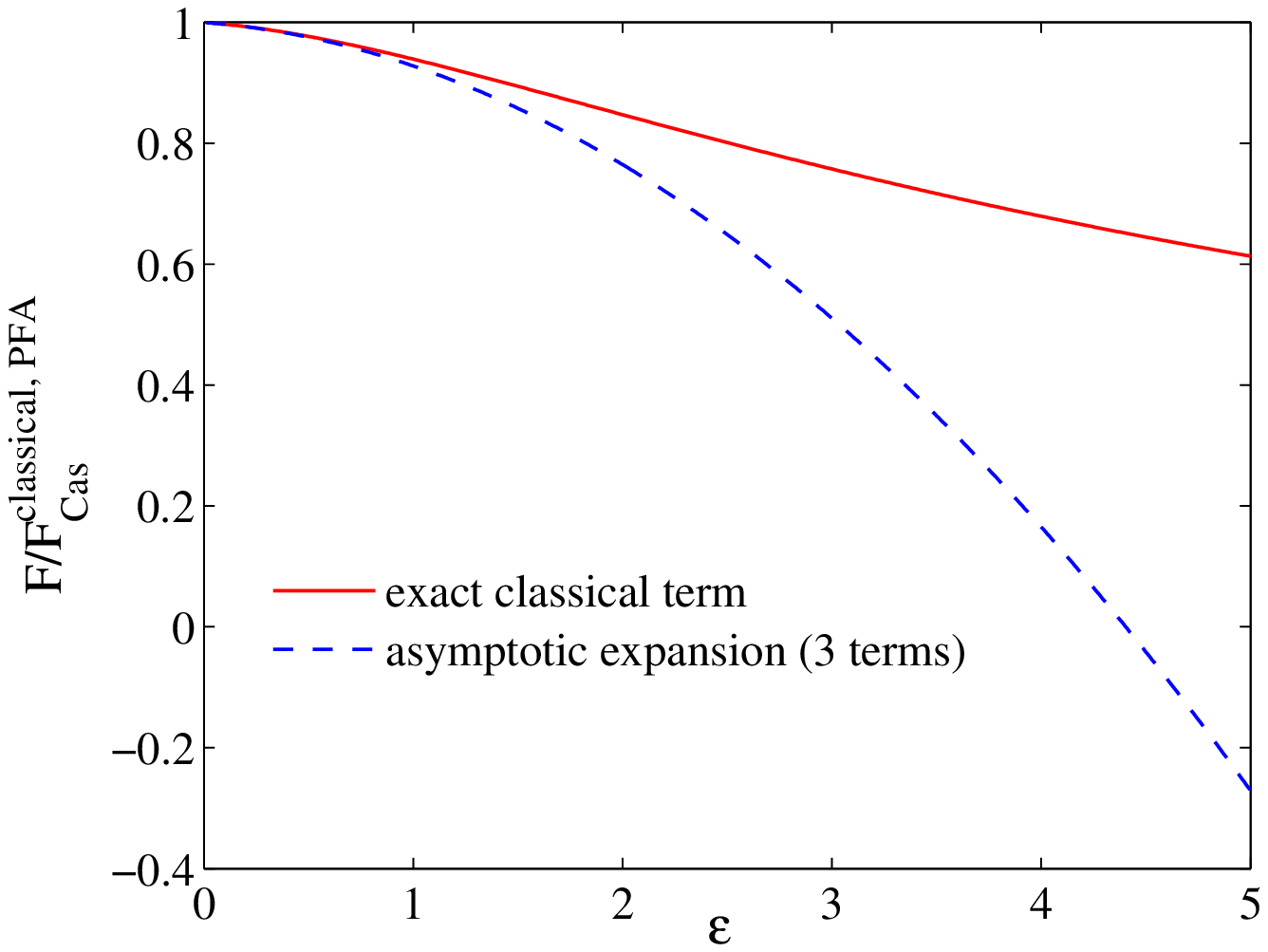} \caption{\label{f3} The comparison between the exact Casimir interaction force with the   asymptotic expansion \eqref{eq3_27_7} when $D=5$. Both quantities are normalized by the proximity force approximation. }\end{figure}

When $D\geq 6$, we do not need to take into account the term \eqref{eq3_26_13} in \eqref{eq3_26_9} nor the second term and third term in \eqref{eq3_26_8}. Eqs. \eqref{eq3_26_8}, \eqref{eq3_26_9},  \eqref{eq3_26_11}, \eqref{eq3_26_12}  and  \eqref{eq3_27_11} give
\begin{equation}\label{eq3_26_14}\begin{split}
F_{\text{Cas}}^{\text{classical}}=&-\frac{k_BT}{R\sqrt{\vep^2+2\vep}}\frac{2}{(D-2)!}\left\{\frac{\Gamma(D)}{2^D\mu^D}\zeta(D)-\frac{(D-2)(D-3)(D-4)}{24}\frac{\Gamma(D-2)}{2^{D-2}\mu^{D-2}}\zeta(D-2)
\right.\\&\hspace{3cm}\left.+\frac{(D-2)(D-3)(D-4)(D-5)(D-6)(5D-8)}{5760}\frac{\Gamma(D-4)}{2^{D-4}\mu^{D-4}}\zeta(D-4)+\ldots\right\}\\
=&-k_BT\frac{(D-1)\zeta(D)}{2^{\frac{3D-1}{2}}R\vep^{\frac{D+1}{2}}}\left(1-\frac{\vep}{4}+\frac{3}{32}\vep^2+\ldots\right)
\left(1+\frac{D\vep}{12}-\frac{(D-3)(D-4)}{3(D-1)}\frac{\zeta(D-2)}{\zeta(D)}\vep\right.\\&\left. +\frac{D(5D-22)}{1440}\vep^2-\frac{(D-2)(D-3)(D-4)}{36(D-1)}\frac{\zeta(D-2)}{\zeta(D)}\vep^2 +\frac{(D-5)(D-6)(5D-8)}{90(D-1)}\frac{\zeta(D-4)}{\zeta(D)}\vep^2+\ldots\right)\\
=&-k_BT\frac{(D-1)\zeta(D)}{2^{\frac{3D-1}{2}}R\vep^{\frac{D+1}{2}}}
\left(1+\frac{(D-3)}{12}\vep-\frac{(D-3)(D-4)}{3(D-1)}\frac{\zeta(D-2)}{\zeta(D)}\vep\right.\\&\left. +\frac{(D-5)(5D-27)}{1440}\vep^2-\frac{(D-3)(D-4)(D-5)}{36(D-1)}\frac{\zeta(D-2)}{\zeta(D)}\vep^2 +\frac{(D-5)(D-6)(5D-8)}{90(D-1)}\frac{\zeta(D-4)}{\zeta(D)}\vep^2+\ldots\right).
\end{split}\end{equation}

\begin{figure}[h]
\epsfxsize=0.5\linewidth \epsffile{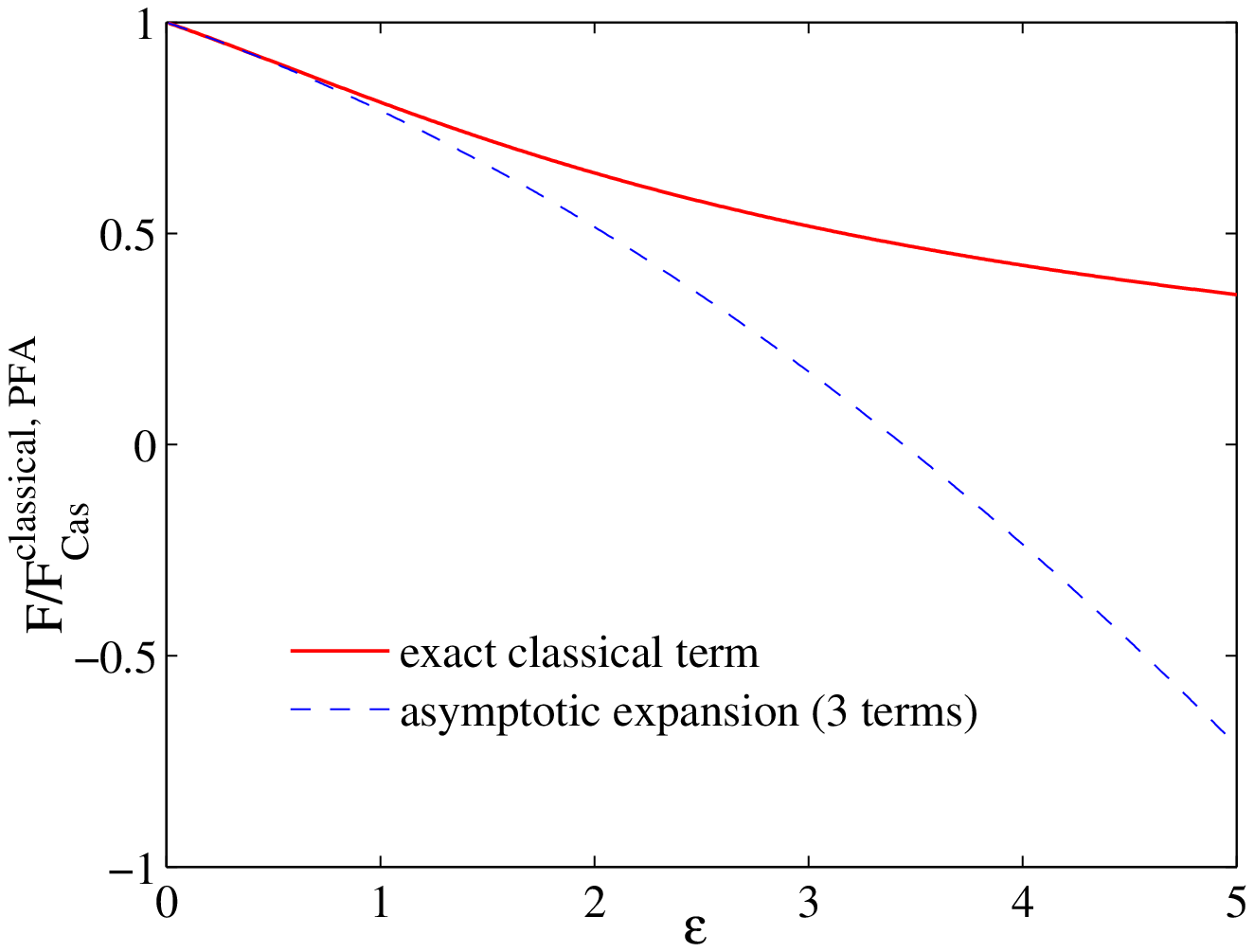} \caption{\label{f4} The comparison between the exact Casimir interaction force with the   asymptotic expansion \eqref{eq3_26_14} when $D=6$. Both quantities are normalized by the proximity force approximation. }\end{figure}

\begin{figure}[h]
\epsfxsize=0.5\linewidth \epsffile{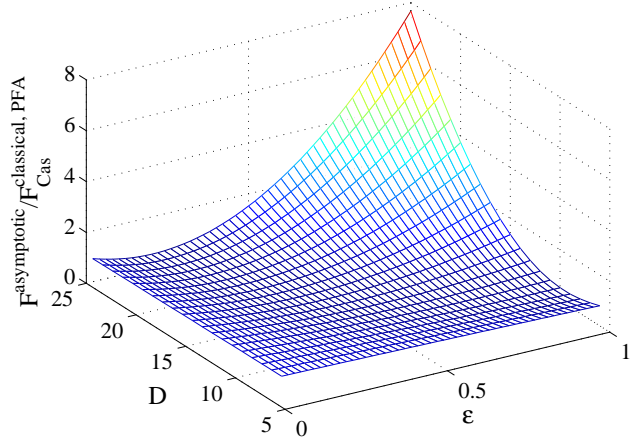} \caption{\label{f5} Dependence of the  asymptotic expansion \eqref{eq3_26_14} on $\vep$ and $D$ for $0\leq \vep\leq 1$ and   $6\leq D\leq 25$. The asymptotic expansion is normalized by the proximity force approximation. }\end{figure}

In Figs. \ref{f1}, \ref{f2}, \ref{f3} and \ref{f4}, we compare the exact classical Casimir interaction force to the three-term asymptotic expansions derived in \eqref{eq3_27_5}, \eqref{eq3_27_6}, \eqref{eq3_27_7} and \eqref{eq3_26_14} when $D=3, 4, 5, 6$. Both quantities are normalized by the proximity force approximation. It is observed that when $\vep=1$, there is a considerable amount of correction to the proximity force approximation, but the three-term asymptotic expansion still give quite good approximation to the exact classical Casimir term. However, the three-term approximation will breakdown when $\vep$ is larger.

In Fig. \ref{f5}, we plot the dependence of the three term asymptotic expansion \eqref{eq3_26_14}, normalized by the proximity force approximation, on the normalized distance $\vep$ and dimension $D$. It is observed that the correction to the proximity force approximation becomes larger when $D$ is larger.
In fact, from \eqref{eq3_26_14}, we find that when $D$ is large,
\begin{equation}\begin{split}
F_{\text{Cas}}^{\text{classical}}\sim &F_{\text{Cas}}^{\text{classical, PFA}}\left(1-\frac{D}{4}\vep+\frac{D^2}{32}\vep^2+\ldots\right).
\end{split}
\end{equation}This shows that the proximity force approximation becomes less accurate in spacetime with higher dimensions.

\section{Conclusion}

In this work, we have computed the high temperature limit for the Casimir free interaction energy between a Dirichlet sphere and a Dirichlet plate in $(D+1)$-dimensional Minkowski spacetime. This high temperature limit is known as the classical term since it does not depend on the Planck constant. It comes from the term with Matsubara frequency zero in the functional representation of the Casimir free energy and can be expressed as a weighted sum of logarithms of  determinants. We derive two alternative exact expressions for this classical term. First we express the logarithms of the determinants as  sums of the logarithms of the eigenvalues. We then use the generalized Abel-Plana summation formula to rewrite this sum so that one can   deduce the small separation asymptotic behaviors of the classical interaction force. The first three terms of the small separation asymptotic expansion are derived explicitly. As expected, the leading term agrees with the proximity force approximation and is proportional to $d^{-\frac{D+1}{2}}$, where $d$ is the distance between the sphere and the plate. The dimension dependence of the next two terms in the expansion are studied and it is found that in higher dimensions, proximity force approximation becomes less accurate.

\begin{acknowledgments}\noindent
  This work is supported by the Ministry of Higher Education of Malaysia  under   FRGS grant FRGS/1/2013/ST02/UNIM/02/2.
\end{acknowledgments}

\end{document}